\documentclass[12pt,preprint]{aastex}
\shorttitle{Cutoff-Free Torsional Wave Propagation}
\shortauthors{Musielak et al.}
\begin{document}
\title{Cutoff-Free Propagation of Torsional Alfv\'en Waves Along 
Thin Magnetic Flux Tubes }
\author{Z.E. Musielak\altaffilmark{1,2}, S. Routh\altaffilmark{1} 
and R. Hammer\altaffilmark{2}}
\altaffiltext{1}{Department of Physics, University of Texas at
                 Arlington, Arlington, TX  76019,\ \ USA; 
                 zmusielak@uta.edu; sxr9573@uta.edu}
\altaffiltext{2}{Kiepenheuer-Institut f\"ur Sonnenphysik, 
                 Sch\"oneckstr. 6, 79104 Freiburg,\ \ Germany;
                 hammer@kis.uni-freiburg.de}                 
\begin{abstract}
Propagation of torsional Alfv\'en waves along magnetic flux tubes 
has been extensively studied for many years but no conclusive results 
regarding the existence of a cutoff frequency for these waves have 
been obtained. The main purpose of this paper is to derive new wave 
equations that describe the propagation of linear torsional Alfv\'en 
waves along thin and isothermal magnetic flux tubes, and use these 
wave equations to demonstrate that the torsional wave propagation 
is not affected by any cutoff frequency. It is also shown that this 
cutoff-free propagation is independent of different choices of the 
coordinate systems and wave variables adopted in the previous studies. 
A brief discussion of implications of this cutoff-free propagation of 
torsional tube waves on theories of wave heating of the solar and stellar 
atmospheres is also given.   
\end{abstract}

\section{Introduction}
%%\hbox to \hsize{%
%% {\vbox to 0pt{\vskip-16cm% <= adjust here!
%%     \nointerlineskip\frenchspacing\small%
%%     \hfill\hbox{\it ApJ, in press, to appear in Vol 660 No 1 (May 1, 2007)}%
%%     \vss}}}%
Direct measurements show that solar magnetic fields outside sunspots 
are concentrated into flux-tube structures located primarily at the 
boundaries of supergranules (e.g., Solanki 1993). Similar magnetic  
structures are likely to exist in late-type stars (e.g., Saar 1996, 
1998). The fundamental modes supported by solar and stellar magnetic
flux tubes can be classified as sausage, kink, torsional Alfv\'en 
and fluting modes (for details, see excellent reviews by Hollweg 1990, 
Roberts 1991, and Roberts \& Ulmschneider 1997, and references therein). 
The role played by these waves in the heating of different parts of the 
solar and stellar atmospheres was discussed by Narain \& Ulmschneider 
(1996) and Ulmschneider \& Musielak (2003). The energy carried by 
sausage and kink tube waves was used as the input to the theoretical 
models of stellar chromospheres constructed by Cuntz et al.\ (1999) 
and Fawzy et al.\ (2002a, b).

In many theoretical studies of tube waves, it is assumed that these 
waves are linear and they propagate along vertically oriented and 
thin flux tubes. Under these conditions, the waves do not interact 
with each other, so they can be investigated independently. Important 
studies of sausage tube waves were performed by Defouw (1976), who 
derived the wave equation for these waves and demonstrated that there 
is a cutoff frequency, which restricts the wave propagation to only 
those frequencies that are higher than the cutoff (see also Webb \& 
Roberts 1979; Rae \& Roberts 1982; Edwin \& Roberts 1983; Musielak, 
Rosner, \& Ulmschneider 1987). Similar studies of kink tube waves 
were performed by Spruit (1981, 1982), who derived both the wave 
equation and the cutoff frequency for these waves (see also Musielak 
\& Ulmschneider 2001).  

Propagation of torsional Alfv\'en waves along solar and stellar magnetic 
flux tubes was extensively studied in the literature (e.g., Parker 1979;
Priest 1982, 1990; Edwin \& Roberts 1983; Hollweg 1985; 1990; Poedts, 
Hermans, \& Goossens 1985; Ferriz-Mas, Sch\"ussler, \& Anton, 1989; 
Roberts 1991; Ferriz-Mas \& Sch\"ussler 1994; Roberts \& Ulmschneider 
1997; Hasan et al.\ 2003; Noble, Musielak, \& Ulmschneider 2003). Two 
different approaches were considered and different sets of wave variables 
were used. In the first approach, the propagation of the waves was described 
in a global coordinate system (e.g., Ferriz-Mas et al.\ 1989; Noble et al.\ 
2003), while in the second approach a local coordinate system was used 
(Hollweg 1978, 1981, 1992). The momentum and induction equations derived 
by Ferriz-Mas et al.\ (1989) were adopted by Ploner \& Solanki (1999) in 
their studies of the influence of torsional tube waves on spectral lines 
formed in the solar atmosphere. In numerical studies of torsional tube 
waves performed by Kudoh \& Shibata (1999) and Saito, Kudoh, \& Shibata 
(2001), the basic equations originally derived by Hollweg were extended 
to more than one dimension and nonlinear terms were included. 

The specific problem of the existence or non-existence of a cutoff 
frequency for torsional Alfv\'en waves propagating along thin and 
isothermal magnetic flux tubes has not been discussed in the 
literature. An exception is the paper by Noble et al.\ (2003), who 
studied the generation rate of torsional tube waves in the solar 
convection zone and introduced the cutoff frequency, defined as the 
ratio of the Alfv\'en velocity to four times the pressure (or density) 
scale height, for these waves. In this paper, we revisit the problem 
by deriving new wave equations that describe the propagation of 
torsional tube waves and demonstrating that this propagation is 
cutoff-free. We also show that the cutoff-free propagation is 
independent of different choices of wave variables and coordinate 
systems used by Ferriz-Mas et al.\ (1989) and Hollweg (1978, 1981,
1992).  

The paper is organized as follows. The momentum and induction 
equations derived in a global coordinate system are presented in \S 2. 
In \S 3, new wave equations for torsional Alfv\'en waves propagating 
along thin and isothermal magnetic flux tubes are obtained and it is 
shown that the wave propagation is not affected by any cutoff frequency. 
The fact that this cutoff-free propagation is independent of different 
choices of wave variables and coordinate systems is demonstrated in \S 4. 
A brief discussion of our results is given in \S 5, and conclusions are 
presented in \S 6.      

\section {Basic equations}

We consider an isolated and vertically oriented magnetic flux tube 
that is embedded in a magnetic field-free, compressible and isothermal 
medium. The tube has a circular cross-section and is in temperature 
equilibrium with the external medium. Let us introduce a global 
cylindrical coordinate system ($r, \phi, z$), with $z$ being the 
tube axis, and describe the background medium inside the tube by 
the gas density $\rho_0 = \rho_0 (r,z)$, the gas pressure $p_0 = 
p_0 (r,z)$ and the magnetic field $\vec B_0 = B_{or} (r,z) \hat r 
+ B_{oz} (r,z) \hat z$. The physical properties of the external 
medium are determined by $\rho_e = \rho_e (r,z)$, $p_e = p_e (r,z)$ 
and $\vec B_e = 0$. Moreover, we also have $T_0 = T_e$ = const. 

To describe torsional Alfv\'en waves, we introduce $\vec v = v_{\phi} 
(r,z,t) \hat \phi$ and $\vec b = b_{\phi} (r,z,t) \hat \phi$, and 
assume that the waves are linear and purely incompressible, which 
means that both the perturbed density $\rho$ and pressure $p$ can 
be neglected. As a result of these assumptions, the propagation of 
the waves is fully described by the momentum and induction equations.
The $\phi$-component of the momentum equation can be written in 
the following form:
     \begin{equation}
{{\partial} \over {\partial t}} \left ( {v_{\phi} \over {r} } \right ) - 
{1 \over {4 \pi \rho_0 r^2}} \left [ B_{0r} {{\partial} \over {\partial r}} 
+ B_{0z} {{\partial} \over {\partial z}} \right ] ( r b_{\phi} ) = 0\ ,
     \label{momenteq1}
     \end{equation}
\noindent
and the $\phi$-component of the induction equation becomes
     \begin{equation}
{{\partial} \over {\partial t}} ( r b_{\phi} ) - r^2 \left [ B_{0r} 
{{\partial} \over {\partial r}} + B_{0z} {{\partial} \over {\partial z}} 
\right ] \left ( {v_{\phi} \over {r} } \right ) = 0\ .
     \label{inducteq1}
     \end{equation}

\noindent
The derived momentum and induction equations are our basic equations 
for all the results derived and discussed in this paper.

\section{Propagation along thin magnetic flux tubes}

\subsection{The thin flux tube approximation}
 
Solar magnetic flux tubes are considered to be thin if their magnetic 
field is horizontally uniform, which means that at a given height all 
magnetic field lines have the same physical properties (e.g., Priest 
1982). The essence of the so-called thin flux tube approximation 
(Roberts \& Webb 1978, 1979; Spruit 1981, 1982; Priest 1982; Hollweg 
1985; Ferriz-Mas, Sch\"ussler, \& Anton, 1989; Ferriz-Mas \& Sch\"ussler 
1994; Musielak et al.\ 1995; Roberts \& Ulmschneider 1997; Hasan et al.\ 
2003) is that radial expansions around the axis of symmetry can be 
truncated at a low order.  For the radial component of the magnetic 
field the leading term is of first order,  
     \begin{equation}
B_{0r} (r,z) = B_{0r} (r,z) \vert_{r=0} + r \left [ {{\partial B_{0r} 
(r,z)} \over {\partial r}} \right ]_{r = 0} +\ ...\ ,
     \label{thineq1}
     \end{equation}

\noindent
since $B_{0r} (r=0, z) = 0$ at the symmetry axis (cf.\ Ferriz Mas
\& Sch\"ussler 1989). Away from the tube axis $B_{0r} (r,z)$ can be 
expressed in terms of $r$ and $B_{0z} (z)$. Using the solenoidal 
condition $\nabla \cdot \vec B_0 = 0$, we obtain 
     \begin{equation}
B_{0r} (r,z) = - {r \over 2} B_{0z}^{\prime} (z)\ ,
     \label{thineq2}
     \end{equation}

\noindent
where $B_{0z}^{\prime} = d B_{0z} / d z$. 

The thin flux tube approximation also requires that $\rho_0 = \rho_0 (z)$, 
$p_0 = p_0 (z)$, $B_{0z} = B_{0z} (z)$, $\rho_e = \rho_e (z)$, $p_e = p_e 
(z)$ and $T_0 = T_e$ = const. In addition, the horizontal pressure balance 
must be satisfied, $p_0 + B_{0z}^2 / 8 \pi = p_e$ at $r = R_t$, where $R_t$ 
is the tube radius. The increase of $R_t$ with height is determined by the 
conservation of the magnetic flux $\pi R_t^2 B_{0z}$ = const. As a result 
of the above assumptions, the Alfv\`en velocity $c_A = B_{0z} / \sqrt{4 \pi 
\rho_0}$ remains constant along the entire length of a thin and isothermal 
magnetic flux tube, and $B_{0z}^{\prime} = - B_{0z} / 2H$, where the pressure 
(density) scale height $H$ is also constant. For reasons explained in the next 
subsection, the wave variables $v_{\phi}$ and $b_{\phi}$ are considered here 
to be functions of time and both spatial coordinates $r$ and $z$.

\subsection{Wave equations}

Using Eq.\ (\ref{thineq2}) and taking into account the fact that the 
variables $r$ and $z$ are independent in the global coordinate system, 
we write Eqs.~(\ref{momenteq1}) and (\ref{inducteq1}) as 
     \begin{equation}
{{\partial v_\phi} \over {\partial t}} + {B_{0z}^{\prime} \over {8 \pi 
\rho_0}} \left ( r {{\partial b_\phi} \over {\partial r}} + b_{\phi} 
\right ) - {B_{0z} \over {4 \pi \rho_0}} {{\partial b_\phi} \over 
{\partial z}} = 0\ ,
     \label{momenteq2}
     \end{equation}
and
     \begin{equation}
{{\partial b_\phi} \over {\partial t}} + {B_{0z}^{\prime} \over 2} 
\left ( r {{\partial v_\phi} \over {\partial r}} - v_{\phi} \right ) 
- B_{0z} {{\partial v_\phi} \over {\partial z}} = 0\ .
     \label{inducteq2}
     \end{equation}

We combine the above equations and derive the wave equations for the wave 
variables $v_{\phi} (r,z,t)$ and $b_{\phi} (r,z,t)$
\[
{{\partial^2 v_{\phi}} \over {\partial t^2}} - c_A^2 
{{\partial^2 v_{\phi}} \over {\partial z^2}} + {c_A^2 \over {2 H}}
{{\partial v_{\phi}} \over {\partial z}} - {c_A^2 \over {16 H^2}} 
v_{\phi} 
\]
     \begin{equation} 
\hskip0.25in - c_A^2 \left ( {r \over {4 H}} \right ) \left [ \left ( 
{r \over {4 H}} \right ) {{\partial^2 v_{\phi}} \over {\partial r^2}} + 
2 {{\partial^2 v_{\phi}} \over {\partial r \partial z}} - {1 \over {4 H}} 
{{\partial v_{\phi}} \over {\partial r}} \right ]
= 0\ ,
     \label{waveq1}
     \end{equation}
and
\[
{{\partial^2 b_{\phi}} \over {\partial t^2}} - c_A^2 
{{\partial^2 b_{\phi}} \over {\partial z^2}} - {c_A^2 \over {2 H}}
{{\partial b_{\phi}} \over {\partial z}} - {c_A^2 \over {16 H^2}} 
b_{\phi} 
\]
     \begin{equation} 
\hskip0.25in - c_A^2 \left ( {r \over {4 H}} \right ) \left [ \left ( 
{r \over {4 H}} \right ) {{\partial^2 b_{\phi}} \over {\partial r^2}} 
+ 2 {{\partial^2 b_{\phi}} \over {\partial r \partial z}} + {3 \over {4 H}} 
{{\partial b_{\phi}} \over {\partial r}} \right ] = 0\ .
     \label{waveq2}
     \end{equation}

\noindent
The derived wave equations show that the value of the parameter $(r / 4 H)$ 
determines the contributions of the $r$-dependence of the wave variables 
$v_{\phi}$ and $b_{\phi}$ to the propagation of torsional tube waves. 
For wide flux tubes, the contributions are important, however, for very 
thin flux tubes with $(r / 4 H) < < 1$, the contributions become negligible. 
It must be noted that the limit $(r / 4 H) \rightarrow 0$ is not allowed 
because $v_{\phi} (r,z,t)\vert_{r=0} = 0$ and $b_{\phi} (r,z,t)\vert_{r=0} 
= 0$. From a physical point of view, this means that a flux tube reduced 
to a single magnetic field line cannot support torsional waves.   

Since the derived wave equations have different forms for $v_{\phi}$ and 
$b_{\phi}$, the wave variables behave differently. In the following, we 
transform these wave equations to new variables that obey the same wave 
equations.

\subsection{Transformed wave equations} 

Using the transformations $v_{\phi} (r,z,t) = v (r,z,t) \rho^{-1/4}$ 
and $b_{\phi} (r,z,t) = b (r,z,t) \rho^{1/4}$ (see Musielak et al.\ 1995; 
Musielak \& Ulmschneider 2001; Noble et al.\ 2003), we obtain
     \begin{equation}
{{\partial^2 v} \over {\partial t^2}} - c_A^2 {{\partial^2 v} 
\over {\partial z^2}} - c_A^2 \left ( {r \over {4 H}} \right ) 
\left [ \left ( {r \over {4 H}} \right ) {{\partial^2 v} \over 
{\partial r^2}} + 2 {{\partial^2 v} \over {\partial r \partial z}} 
+ {1 \over {4 H}} {{\partial v} \over {\partial r}} \right ] = 0\ ,
     \label{waveq3}
     \end{equation}
and
     \begin{equation}
{{\partial^2 b} \over {\partial t^2}} - c_A^2 {{\partial^2 b}
\over {\partial z^2}} - c_A^2 \left ( {r \over {4 H}} \right ) 
\left [ \left ( {r \over {4 H}} \right ) {{\partial^2 b} \over 
{\partial r^2}} + 2 {{\partial^2 b} \over {\partial r \partial z}} 
+ {1 \over {4 H}} {{\partial b} \over {\partial r}} \right ] = 0\ .
     \label{waveq4}
     \end{equation}

\noindent
Clearly, the behavior of the transformed wave variables $v$ and $b$ 
is identical.

To remove the first-order derivatives from the above equations, we use the 
transformation $d \zeta = (4 H / r) d r$, which gives 
    \begin{equation}
\left ( {{\partial^2} \over {\partial t^2}} - c_A^2 {{\partial^2} \over 
{\partial z^2}} - c_A^2 {{\partial^2} \over {\partial \zeta^2}} - 2 c_A^2 
{{\partial^2} \over {\partial z \partial \zeta}} \right ) 
[ v (\zeta,z,t); b (\zeta,z,t) ] = 0\ ,
     \label{waveq5}
     \end{equation}

\noindent
where $\zeta = 4 H\  \ln \vert r \vert$. This is the most general equation 
that describes the propagation of torsional waves along thin and isothermal 
magnetic flux tubes. The equation shows that there is no cutoff frequency 
for torsional tube waves (see Sec.\ 3.4). 

\subsection{Dispersion relation}

Since all the coefficients in Eq.~(\ref{waveq5}) are constant, we make 
Fourier transforms in time and space, and derive the following dispersion
relation
     \begin{equation}
\omega^2 = (k_z^2 + 2 k_z k_{\zeta} + k_{\zeta}^2) c_A^2\ ,
     \label{dispeq1}
     \end{equation}

\noindent
where $\omega$ is the wave frequency and $k_z$ and $k_{\zeta}$ are the 
$z$ and $\zeta$ components of the wave vector $\vec k$, respectively.  
Note that the same dispersion relation is obtained for each wave 
variable. 

Let us define $\kappa = k_z + k_{\zeta}$ and write  
     \begin{equation}
\omega^2 = \kappa^2 c_A^2\ ,
     \label{dispeq2}
     \end{equation}

\noindent
which shows that the propagation of linear torsional Alfv\'en waves 
along thin and isothermal magnetic tube waves is not affected by any 
cutoff frequency.   

\section{Other approaches}

To demonstrate that the propagation of torsional tube waves is cutoff-free, 
we used the global coordinate system and the original wave variables 
$v_{\phi}$ and $b_{\phi}$, which were transformed to the new variables 
$v$ and $b$. Two different approaches were developed by Ferriz-Mas et al.\ 
(1989), who adopted the same coordinate system but used different wave 
variables, and by Hollweg (1978, 1981, 1992), who chose a local coordinate
system and introduced different wave variables (see also Edwin \& Roberts 
1983 and Poedts et al.\ 1985). In addition, Noble et al.\ (2003) considered 
the global coordinate system and used the wave variables $v_{\phi}$ and 
$b_{\phi}$. However, their assumption that $B_{0r} = 0$ was inconsistent 
with the solenoidal condition, which makes their claim of the existence 
of the cutoff frequency for torsional Alfve\'en waves invalid. In the 
following, we demonstrate that the momentum and induction equations 
derived by Ferriz-Mas et al.\ and Hollweg lead to the same results as
those found in Sec.\ 3 of this paper. 

\subsection{Global coordinate system and different wave variables}

In their work on propagation of waves along thin magnetic flux tubes, 
Ferriz-Mas et al.\ (1989) used the global coordinate system and derived 
the first and second-order equations that describe the propagation of 
sausage, kink and torsional Alfv\'en tube waves. In their approach,
each wave variable is expanded in a Taylor series and, specifically 
for torsional tube waves, the new variables $v_{\phi1}$ and $b_{\phi1}$, 
which represent the first order expansion in the series, are introduced. 
These variables are given by 
\begin{equation}
v_{\phi1} (z,t) = {{\partial v_{\phi}} \over {\partial r}}\vert_{r=0} 
\hskip0.3in {\rm and }\hskip0.3in
b_{\phi1} (z,t) = {{\partial b_{\phi}} \over {\partial r}}\vert_{r=0}\ . 
\label{fsaeq1}
\end{equation}

Since $v_{\phi} (r,z,t)\vert_{r=0} = 0$ and $b_{\phi} (r,z,t)\vert_{r=0} 
= 0$ (see Sec.\ 3.2), we may use Eq.\ (\ref{fsaeq1}) to write $v_{\phi} 
(r,z,t) = r v_{\phi1} (z,t)$ and $b_{\phi} (r,z,t) = r b_{\phi1}
(z,t)$ in first order. Substituting these new variables into Eqs.\ 
(\ref{momenteq2}) and (\ref{inducteq2}), we obtain
     \begin{equation}
{{\partial v_{\phi1}} \over {\partial t}} + {1 \over {4 \pi \rho_0}} 
\left ( B_{0z}^{\prime} b_{\phi1} - B_{0z} {{\partial b_{\phi1}} 
\over {\partial z}} \right ) = 0\ ,
     \label{fsaeq2}
     \end{equation}
and
     \begin{equation}
{{\partial b_{\phi1}} \over {\partial t}} - B_{0z} {{\partial v_{\phi1}} 
\over {\partial z}} = 0\ ,
     \label{fsaeq3}
     \end{equation}

\noindent
which are the same equations as those obtained by Ferriz-Mas et al.\ (1989, 
see their Eqs. 14 and 16; note that our $B_{r0}$ corresponds to their
$r B_{r1}$, and $v_z$ and $v_r$ vanish in our case). Despite the fact that 
$v_{\phi1}$ and $b_{\phi1}$ depend solely on $z$ and $t$, the above equations 
are not valid at the tube axis (see discussion above).
 
The wave equations resulting from the above momentum and induction equations 
become
    \begin{equation}
{{\partial^2 v_{\phi1}} \over {\partial t^2}} - c_A^2 {{\partial^2 v_{\phi1}} 
\over {\partial z^2}} = 0\ ,
     \label{fsaeq4}
     \end{equation}
and
    \begin{equation}
{{\partial^2 b_{\phi1}} \over {\partial t^2}} - c_A^2 {{\partial^2 b_{\phi1}} 
\over {\partial z^2}} - {c_A^2 \over H} {{\partial b_{\phi1}} \over {\partial z}}
- {c_A^2 \over {4 H^2}} b_{\phi1} = 0\ .
     \label{fsaeq5}
     \end{equation}

\noindent
Clearly, the derived wave equations have different forms, which implies that 
the wave variables $v_{\phi1}$ and $b_{\phi1}$ behave differently. To remove
the first-order derivative from Eq.\ (\ref{fsaeq5}), we use the transformation
$b_{\phi1} (z,t) = \tilde b_{\phi1} (z,t) \rho^{1/2}$, and obtain
    \begin{equation}
{{\partial^2 \tilde b_{\phi1}} \over {\partial t^2}} - c_A^2 
{{\partial^2 \tilde b_{\phi1}} \over {\partial z^2}} = 0\ .
     \label{fsaeq6}
     \end{equation}

\noindent
Hence, the behavior of the wave variables $v_{\phi}$ and $\tilde b_{\phi1}$ 
is the same and there is no cutoff frequency that affects the wave propagation.
This is an important result as it shows that the non-existence of a cutoff
frequency for torsional tube waves is independent of the choice of the 
wave variables. 

\subsection{Local coordinate system and different wave variables}

Propagation of torsional Alfve\'n waves along magnetic flux tubes can also 
be described by using a local orthogonal curvilinear coordinate system 
($\xi$, $\theta$, $s$), with $s$ being the length measured along a magnetic 
field line, $\theta$ the azimuthal angle about the axis of symmetry, and 
$\xi$ a coordinate in the direction $\hat{\xi} = \hat{\theta}\times\hat{s}$. 
In this case, $\vec B_0 = B_{0s} (s) \hat s$, $B_{0\xi} = 0$ and $B_{0\theta} 
= 0$. We also have $\vec v = v_{\theta} (s,t) \hat{\theta}$, $\vec b = 
b_{\theta} (s,t) \hat{\theta}$ and $R = R(s)$, where $R$ represents the 
distance from the magnetic field line to the tube axis. This approach was 
first considered by Hollweg (1978), who also applied it to solar magnetic 
flux tubes (see Hollweg 1981, 1992).

Following Hollweg, Jackson, \& Galloway (1982), the curvilinear scale
factors are $h_{\phi} = R$ and $h_s = 1$, and we determine $h_{\xi}$ from 
the condition $h_{\xi}\ R\ B_{0s}$ = const, which results from $\nabla 
\cdot \vec B_0 = 0$. To conserve the magnetic flux, we must choose $h_{\xi} 
= R$. Using these scale factors, the explicit form of the momentum and 
induction equations is
     \begin{equation}
{{\partial} \over {\partial t}} \left ( {v_{\theta} \over {R} } 
\right ) - {B_{0s} \over {4 \pi \rho_0 R^2}} {{\partial} \over 
{\partial s}} ( R b_{\theta} ) = 0\ ,
     \label{Holleq1}
     \end{equation}
and 
     \begin{equation}
{{\partial} \over {\partial t}} ( R b_{\theta} ) - R^2 B_{0s} 
{{\partial} \over {\partial s}} \left ( {v_{\theta} \over {R} } 
\right ) = 0\ .
     \label{Holleq2}
     \end{equation}

It is easy to see that the magnetic field $\vec B_0 (s)$ in the local 
coordinate system can be described in the global coordinate system 
(see Eqs.~\ref{momenteq1} and \ref{inducteq1}) by the $B_{0r} (r,z)$ 
and $B_{0z} (r,z)$ field components. This means that the following  
relation must hold between the spatial operators in these two 
coordinate systems
     \begin{equation}
B_{0r} {{\partial} \over {\partial r}} + B_{0z} {{\partial} 
\over {\partial z}} = B_{0s} {{\partial} \over {\partial s}}\ .
     \label{Holleq3}
     \end{equation}

\noindent
This relation is consistent with the fact that $\vec B_0 \cdot 
\nabla$ must be the same in the global (the LHS of Eq.\ 
\ref{Holleq3}) and local (the RHS of Eq.\ \ref{Holleq3}) 
coordinate systems. Moreover, $v_{\phi}$ and $v_{\theta}$ are 
also related, as the former can be treated as a projection of
the latter on the $\phi$-axis of the global coordinate system.
Obviously, the same is true for the wave variables $b_{\phi}$ 
and $b_{\theta}$. 

We follow Hollweg (1978, 1981) and introduce the new variables 
$x = v_{\theta} / R$ and $y = R b_{\theta}$. The wave equations 
for these variables are
     \begin{equation}
{{\partial^2 x} \over {\partial t^2}} - c_A^2 {{\partial^2 x} 
\over {\partial s^2}} = 0\ ,
     \label{Holleq4_01}
     \end{equation}
and 
     \begin{equation}
{{\partial^2 y} \over {\partial t^2}} - {\partial \over {\partial s}} 
\left ( c_A^2 {{\partial y} \over {\partial s}} \right ) = 0\ ,
     \label{Holleq5}
     \end{equation}

\noindent
where in general $c_A = c_A (s)$. However, for thin magnetic flux tubes
$c_A$ = const (see Sec.\ 3.1) and the wave equations become
     \begin{equation}
\left ( {{\partial^2} \over {\partial t^2}} - c_A^2 {{\partial^2} \over 
{\partial s^2}} \right ) [x (s,t); y (s,t)] = 0\ .
     \label{Holleq4_02}
     \end{equation}

\noindent
Again, no cutoff frequency exists.  It is a significant (but expected) 
result that this non-existence of a cutoff frequency for torsional tube 
waves is independent of the choice of the cooordinate system and the 
wave variables.

\section{Discussion}

We considered the propagation of linear torsional Alfv\'en waves along 
thin and isothermal magnetic flux tubes using the global coordinate 
system, and derived new wave equations describing this propagation. The 
derived wave equations were then used to demonstrate that no cutoff 
frequency exists for these waves, which means that torsional waves of 
any frequency are freely propagating along the tubes. We also showed 
that this result is independent of different choices of the coordinate 
systems and wave variables adopted by Ferriz-Mas et al.\ (1989) and 
Hollweg (1978, 1981, 1992).  

As first shown by Defouw (1976) for sausage tube waves and by Spruit 
(1981) for kink tube waves, the propagation of both modes is affected 
by their corresponding cutoff frequencies. With their cutoff-free 
propagation, torsional Alfv\'en waves seem to be exceptional among the 
tube modes. In general, the existence of cutoff frequencies is caused 
by either gravity or gradients of the characteristic wave velocities, 
which result from an inhomogeneity of the background medium. In the 
cases discussed in this paper, the characteristic wave velocities are 
constant for all tube modes because of the thin flux tube approximation. 
Hence, it is gravity which leads to the origin of the cutoff through 
either stratification (sausage tube waves) or buoyancy force (kink tube 
waves).

The fact that stratification leads to a cutoff frequency for acoustic 
waves propagating in a stratified and isothermal medium was first 
demonstrated by Lamb (1908, 1911). Since sausage tube waves are 
essentially acoustic waves guided by the tube magnetic field, and since 
they propagate in a stratified and isothermal medium inside the tube, 
it is stratification of the background medium that is responsible for 
the existence of the cutoff frequency for these waves. 

The nature of kink tube waves is significantly different than 
sausage tube waves and yet it is again gravity that is responsible
for the existence of the cutoff frequency for these waves. The main 
reason is that magnetic tension and buoyancy are the restoring forces
for kink tube waves, and that the buoyancy force through its dependence 
on gravity leads to the cutoff frequency (e.g., Spruit 1982; Hollweg 
1985), which is lower than that for longitudinal tube waves.

Now, despite some similarities between kink and torsional Alfv\'en 
tube waves, the main difference is that magnetic tension is the only 
restoring force for the latter. Since linear torsional tube waves 
have only purely axisymmetric twists in the $\phi$-direction and 
show no pressure fluctuations, the twists are neither coupled 
to the gravitational force nor affected by stratification. As a 
result, no cutoff frequency can exist for linear torsional Alfv\'en 
waves propagating along thin and isothermal magnetic flux tubes.
 
The cutoff-free propagation of torsional tube waves may have important 
implications on theories of wave heating of the solar and stellar 
atmospheres. The theoretical models of stellar chromospheres constructed
by Fawzy et al.\ (2002a,b) are based on the amount of energy carried by
acoustic waves and by sausage and kink tube waves; these waves are 
generated by turbulent motions in the solar and stellar convection zones. 
The models point to a missing amount of heating for stars with high levels 
of activity. It is likely that the energy carried by torsional Alfv\'en 
waves could be used, at least partially, to account for these ``heating 
gaps''. Since there is no cutoff frequency for torsional tube waves, a broad 
spectrum of these waves is expected to be generated in the solar and stellar 
convection zones. The waves of different frequencies of this spectrum may 
transfer energy to different parts of the solar and stellar atmospheres.  
Hence, new studies are required to determine the efficiency of generation 
of torsional tube waves and their dissipation rates.

\section{Conclusions}

We derived new wave equations that describe the propagation of linear 
torsional Alfv\'en waves along thin and isothermal magnetic flux tubes, 
and used them to demonstrate that this propagation is cutoff-free. Our 
study also showed that the result is independent of different choices 
of the coordinate systems and wave variables used by Ferriz-Mas et al.\ 
(1989) and Hollweg (1978, 1981, 1992).

Since the existence of cutoff frequencies for sausage and kink tube waves 
is caused by stratification and buoyancy force, respectively, our results 
clearly show that neither stratification nor buoyancy force affects the 
torsional wave propagation. The physical reason is that magnetic tension 
is the only restoring force for these waves.   

This lack of any cutoff frequency for torsional tube waves implies 
that a broad wave energy spectrum for these waves will be generated
in the solar and stellar convection zones. The energy carried by the
waves of different frequencies of this spectrum may be used to account 
for the ``heating gaps'' discovered by Fawzy at al.\ (2002a,b) and may  
also contribute to the wave heating of different parts of the solar 
and stellar atmospheres.  

\acknowledgements
This work was supported by NSF under grant ATM-0538278 (Z.E.M. 
and S.R.) and NASA under grant NAG8-1889 (Z.E.M. and S.R.). Z.E.M. 
also acknowledges the support of this work by the Alexander von 
Humboldt Foundation. 

\end{document}